\documentstyle[11pt,newpasp,twoside]{article}
\def\@jourvol{220}
\def\cpr@year{2003}
\def\vol@title{Dark Matter in Galaxies}
\def\vol@author{R. Ryder, D.J. Pisano, M. Walker, K. Freeman}

\def\edcomment#1{\iffalse\marginpar{\raggedright\sl#1\/}\else\relax\fi}
\reversemarginpar

\begin{document}
\title{Alternatives to Dark Matter (?)}
 \author{Anthony Aguirre}
\affil{Department of Physics/SCIPP, UC Santa Cruz\\ 1156 High St. 
Santa Cruz, CA 95064; aguirre@scipp.ucsc.edu}

\begin{abstract}
It has long been known that Newtonian dynamics applied to the visible
matter in galaxies and clusters does not correctly describe the
dynamics of those systems. While this is generally taken as evidence
for dark matter it is in principle possible that instead Newtonian
dynamics (and with it General Relativity) breaks down in these
systems. Indeed there have been a number of proposals as to how
standard gravitational dynamics might be modified so as to correctly
explain galactic dynamics without dark matter. I will review this
general idea (but focus on ``MOdified Newtonian Dynamics'', or ``MOND''), and
discuss a number of ways alternatives to dark matter can be tested and
(in many cases) ruled out.
\end{abstract}

\section{Introduction}
The great majority of astronomers now believe that the universe is
dominated by cold, collisionless, non-baryonic dark matter.  But
despite more than 20 years of intense effort, no {\em
non-gravitational} evidence for dark matter has ever been found: no
direct detection of dark matter, no annihilation radiation from it, no
evidence from reactor experiments supporting the physics (beyond the
standard model) upon which dark matter candidates are based.  We know
nothing about dark matter, except for the properties that we have
attributed to it, and also that it is not enough: we need to postulate
an even more mysterious ``dark energy" to supplement it.

It therefore seems worth keeping in mind, even now, the possibility
that what we have interpreted as evidence for dark matter is, in
fact, evidence for the breakdown of Einstein's (and Newton's) gravity
on scales of galaxies and cosmology.  And indeed, over the years, a
number of rebellious researchers have proposed modifications of
gravitational dynamics as substitutes for dark matter.  The idea
behind all of these is to increase the strength of gravity on galaxy
scales and above so as to explain (for example) the flat rotation
curves of galaxies using only the visible baryonic matter in them.  To
see how this works in detail, and elucidate some of the key points of
making such a modification of gravity, let us go through an extended
example of an attempt to do away with dark matter.

A first guess would be to add to the normal Newtonian gravitational
acceleration a new term that becomes dominant for large radius $r$:
\begin{equation}
a=M\left({G\over r^2}+g(r)\right),
\end{equation}
where $g(r)$ is a free function that does not fall off as fast as
$1/r^2$.  Now let us require that the rotation curves of galaxies are,
as observed, flat at large $r$.  An obvious way to do this is to set the
acceleration at large radii equal to the centrifugal acceleration for
a constant velocity to obtain:
\begin{equation}
g(r)=A/r.
\end{equation}
the problem with this proposal is that if $A$ is independent of $M$,
then $v_\infty \equiv v(r\rightarrow\infty) \propto \sqrt{M}$,
contradicting the observed Tully-Fisher (TF) relation for spiral galaxies that
$M\propto v_\infty^\alpha$ with $\alpha\approx 3.5-4$.  This
important contradiction rules out many alternative gravity theories
(see Aguirre et al. 2001); nearly any alternative gravity in which the
modification becomes important at a fixed {\em length} scale leads to
the wrong TF relation.

We can, however, repair our candidate theory by setting $A \propto
M^{-1/2}$, which yields flat rotation curves as well as $\alpha=4$.  So
we now have:
\begin{equation}
a={GM\over r^2}+{BM^{1/2}\over r}.
\end{equation}
This theory has a couple of funny features.  First, it is nonlinear,
so doubling the mass of a galaxy will not double the gravitational
acceleration; also Newton's third law is not obeyed if the equation is
written in terms of forces (i.e. momentum is not conserved).  Second,
the modification becomes important at a fixed physical {\em
acceleration} $a_0 = B^2/G$ . As we will see this theory is very
similar to Milgrom's well known MOdified Newtonian Dynamics (MOND),
which is no coincidence: Milgrom used similar arguments to develop
MOND, which is, I suspect, essentially the unique modification of
Newtonian gravity giving {\em asymptotically flat} rotation curves as
well as the correct TF relation.

A fixed acceleration scale appears to be a part of any explanation of
the systematics of galaxy rotation curves.  For example, we can look
at three modified gravity theories from the literature that claim to
provide fits to galaxy rotation curves as well as explain the
TF relation. First, Mannheim's (1993; 1997) ``conformal
gravity" theory has, in the non-relativistic limit of a test particle
near a spherical mass distribution of mass $M$, the acceleration law:
\begin{equation}
a = {GM\over r^2}+BM+a_0,
\end{equation}
where $B$ and $a_0$ are fixed constants.  It turns out that to fit
galaxy rotation curves, $B$ must be fairly small so that the first and
third terms dominate; thus the modification becomes important at a
fixed acceleration scale $a_0$.  Note, however, that rotation curves
in this theory are not asymptotically flat: just as in dark-matter
theory, there is only a range in which they are approximately flat.
Moffat's ``nonsymmetric gravity" theory (see Moffat \& Sokolov 1996)
has a different value of $G$ at small and large radii, but the
transition radius depends on $\sqrt{M}$, as is familiar from our
example theory.

Finally, Milgrom's (1983) MOND theory has the force law that $a=a_N$
for $a_N \gg a_0$ and $a=\sqrt{a_N a_0}$ for $a_N \ll a_0$ (with some
interpolated behavior between these regimes). It thus explicitly
transitions at a fixed acceleration scale and behaves as $\sqrt M/r$
at low accelerations.

\section{MOND, galaxies, and MIFF}

Although all of these theories are fun to think about, I'll focus for
the rest of this talk on MOND, making occasional comments regarding
other modified gravity theories.  For an extensive review of MOND, see
McGaugh \& Sanders (2002).  In brief, MOND as a modification of
gravity (See Bekenstein \& Milgrom 1984) replaces the usual Poisson
equation:
\begin{equation}
\nabla^2\phi=4\pi G\rho \Rightarrow \nabla\cdot[\mu(|\nabla\phi|/a_0)\nabla\phi]=4\pi G\rho,
\end{equation}
where $\mu(x)$ is some interpolating function with the property that
$\mu(x)=x$ for $x \ll 1$ and $\mu(x)=1$ for $x \gg 1$.  Because this can
be derived from a Lagrangian, it ensures conservation of energy,
momentum, and angular momentum.  While the mutual acceleration between
two similar-mass objects is complicated, the acceleration of a test
particle near a satirically symmetric mass distribution is given by
the recipe
\begin{equation}
\mu(|\vec a|/a_0)\vec a=\vec a_N,
\end{equation}
where $a_N$ is again the usual Newtonian acceleration.  This formula
gives the previously described behavior in the high-and
low-acceleration regimes.  An important feature of this formulation,
and one that is essential to any phenomena logically viable
formulation of MOND, is what may be called the ``external field effect".
It can be shown that if system is embedded in an external
gravitational field of Newtonian acceleration $a_{\rm ext}$, then if
the internal (Newtonian) acceleration of the system $a_N$ satisfies
$a_N < a_{\rm ext} < a_0$, then the modified acceleration, rather than
being $\sqrt{a_0 a_N}$, is instead $a_N(a_0/a_{\rm ext})$ in a
coordinate system $(x',y',z')=(x,y,2z)$, where $\hat z \perp \vec
a_{\rm ext}$ and the Newtonian acceleration is computed in the
unprimed coordinates.  That is, the acceleration becomes {\em
Newtonian} (but with a larger effective gravitational constant) and
{\em anisotropic}.  This is an important point to which I will
repeatedly return.

So what were the predictions of this (1984) theory? There were four
principal ones: first, that rotation curves of isolated
galaxies should be asymptotically flat.  Second, that the TF relation
for isolated galaxies would satisfy $M = v^4_\infty/4Ga_0$ exactly.
Third, that the effects of modified gravity would manifest at a
critical surface density, and that if galaxies existed that always
fell below this critical density, they would appear to be dark
matter-dominated everywhere. Fourth, the rotation curves of galaxies
should be calculable given {\em only} the distribution of baryonic
matter, using the MOND formula.

Twenty years later, it must be said that these predictions are holding
up pretty well.  In particular, the fourth prediction that galaxy
rotation curves should be calculable only given their baryonic mass
distributions, has been demonstrated in several studies (see Sanders
\& McGaugh 2001 for a summary) and appears to work to an amazing
degree.  This is not to say that other theories cannot fit these
rotation curves -- for example, Mannheim has shown that he can fit a
number of galaxy rotation curves in detail with his model, and
reasonable fits using disks and dark matter halo profiles can be found
for most of the galaxies -- but MOND accomplishes this with a single
parameter, the baryonic mass-to-light ratio (which is, in fact, almost
a constant when the analysis is done in I-band).  This remarkable
regularity in the properties of galaxies is a phenomenon that, as many
people have pointed out, requires explanation.

Although extremely successful in spiral galaxies, it is somewhat less
clear how well MOND does in other galaxies, e.g., ellipticals.  MOND
is consistent with scaling relations such as the Faber-Jackson or
fundamental plane (Sanders 2000), but somewhat more difficult to test
in ellipticals due to the ambiguity in the anisotropy of the velocity
dispersion.  Nonetheless, there are some interesting recent studies
bearing upon MOND.  For example, the study of Gerhard et al. (2001)
finds that the M/L ratio in ellipticals stars to rise (leading to the
inference of dark matter) at a higher characteristic acceleration than
in spirals, disfavoring a universal critical acceleration $a_0$.  On the other
hand, the recent study of Romanowsky et al. (2003) indicates that
several large ellipticals are {\em deficient} in dark matter, a fact
which apparently finds easy explanation in MOND (Milgrom \& Sanders
2003) because the acceleration regime probed is Newtonian or
near-Newtonian.  On a similar topic, Prada et al. (2003) have recently
measured velocity dispersion curves for dwarf galaxies around massive
field galaxies in the SLOAN survey.  They find a profile that agrees
well with the standard CDM prediction and claim inconsistency with MOND.
However, the external field effect may be important here: at distances
$\ga 50\,$kpc from an $L_*$ spiral galaxy, the internal acceleration
can fall below the typical large-scale flow acceleration, and a
Keplerian velocity falloff -- rather than a flat profile -- would be
expected.

To summarize the status of alternatives to dark matter in galaxies:
first, rather simple considerations lead inexorably to a modification
of gravity similar to MOND, and in particular with a characteristic
acceleration built into the theory.  Second, the formula given by
Milgrom provides an excellent description of the systematics of spiral
galaxy rotation curves.  Milgrom's fitting formula (MIFF) encompasses,
but goes well beyond the TF relation, and could prove to be at least
as useful.  Regardless of the status of MOND as a modification of
gravity, MIFF should be tested (especially in other sorts of galaxies)
and used, without embarrassment, as an excellent phenomenological
relation.

But what about when we leave the realm of galaxies?  If MOND genuinely
constitutes an alternative to dark matter, then it should be able to
reproduce the successfully replace dark matter on cluster,
supercluster, and cosmological scales as well.

\section{MOND and clusters}

It is now well-established that MOND can{\em not} account for the
dynamics of clusters in terms of only the visible (galaxies and X-ray
emitting gas) matter in them.  There are two good ways of seeing
this.  

The first is that in the cores of clusters, strong
gravitational lensing indicates a much larger mass than can be
accounted for by visible stars and gas.  Moreover, the surface density
in cluster cores is sufficiently high that MOND should not apply; thus
we have a system in the Newtonian regime that requires unseen matter
(Milgrom 1999; Sanders 1999)

A second probe of gravitation in clusters is the X-ray emitting gas,
which can to a sufficiently good approximation be considered to be in
hydrostatic equilibrium in cluster's gravitational potential.  Then,
just as a rotation curve can be predicted using the observed mass and
and MOND, so can the temperature profile of the cluster be predicted,
up to one free parameter which is the central cluster temperature.
This was first done in The \& White (1988), where they found a poor
fit unless the MOND constant was significantly larger than that
inferred from galaxies.  However, the observational data was not of
high quality, and the agreement within a factor of $\sim 2$ was
considered ``reasonably good'' by MOND optimists (Sanders 1999).  More
recently, with the advent of higher-quality X-ray data, a more
detailed analysis became possible.  In Aguirre, Schaye \& Quataert
(2001) it was shown that MOND fails to fit the temperature profiles of
several well-observed clusters.  Quantitatively, fitting the observed
profiles within MOND requires additional unobserved matter of $\approx
2-5\times$ the observed baryonic mass within 1\,Mpc, and $\approx
10\times$ the observed mass within 100\,kpc.  Similar results were
obtained in a subsequent larger but less-detailed study by Sanders
(2003).

The necessity of dark matter even given MOND is discouraging at the
very least for a theory formulated to remove the need for dark matter;
but it is not immediately fatal, as it would have been if clusters in
MOND required {\em less} than the observed amount of matter.  Instead,
clusters can be ``saved'' in MOND -- just as galaxies were saved in
Newtonian mechanics -- by postulating extra dark matter.  What might
this be?

It would be difficult to swallow traditional cold non-interacting dark
matter (e.g. WIMPs) alongside MOND, but even if one could, it would
not help: the dark matter in MOND must be unique to clusters.  One
might imagine that there is some baryonic, but cluster-specific form
of dark matter such as warm gas or MACHOs, but I think it is fair to
say none that of the possibilities for this are particularly
attractive, and many can be ruled out.

Probably the most interesting possibility would be neutrinos of mass
$\approx 2\,$eV, an idea adumbrated in Sanders (2003).  These would
aggregate in cluster-scale potentials but not typical galaxies, and
could provide a mass several times that of the baryons.  The required
neutrino mass is near the upper range allowed by current laboratory
experiments, however, and may be tested soon.  This possibility has
the additional advantage that $\approx 2\,$eV neutrinos would
flagrantly violate limits in the standard cosmological model based on
CMB and large-scale structure.  Thus this is an interesting
prediction of MOND that would be extremely difficult to explain in the
standard cosmological model.

\section{MOND as a fundamental theory}

MOND, as formulated for example by Bekenstein \& Milgrom (1984) is a
replacement for Newtonian physics; both fully explain the dynamics of
non-relativistic particles under the gravitation of an arbitrary mass
distribution, but require the fixing of an unaccelerated reference
frame by fiat.  There is, however, no satisfactory generalization of
MOND analogous to General Relativity, and there are in fact ``no-go''
theorems ruling out relativistic MOND theories of various types (e.g.,
Bekenstein \& Sanders 1994; Soussa \& Woodard 2003).  Clever people
with a deep knowledge of physics have tried and failed to relativise
MOND; the conclusion that it therefore cannot be done is often
referred to as ``Bekenstein's theorem'' (though I think this does some
disservice to Milgrom and Sanders).  But this theorem should not be taken too
seriously, as similar theorems also imply that neither quantum gravity, nor
M-theory, nor a realistic model of inflation exist.  I shall not say
more about relativistic MOND, but refer the reader to the review of
Sanders \& McGaugh (2002) and references therein.

In the meantime, MOND simply does not make unambiguous predictions for
phenomena involving relativistic physics such as the cosmic
expansion, the early universe, or gravitational lensing.\footnote{ It
is worth noting that some other modifications of gravity, for example
Mannheim's conformal gravity, {\em do} follow from relativistic
theories, and hence make firm predictions for relativistic phenomena
that can be tested.}  Nevertheless, MOND is not completely mute on
these subjects, as discussed in the next few sections.

\section{MOND and Lensing}

Although there is no rigorous prediction for the dynamics of
relativistic particles such as photons in MOND, a number of arguments
have nevertheless been put forward claiming to test MOND using lensing
data.  In some, a heuristic prescription is applied to generate
lensing predictions in MOND, e.g. that the deflection angle is half
that predicted in the $m\rightarrow 0$ limit of the (MOdified)
Newtonian dynamics. Such arguments, I think, can never rigorously test
MOND, because the assumptions made are not strictly required.

In a more interesting and dangerous (for MOND) class of arguments one
looks for tests that can be applied independent of the details of how
lensing occurs in MOND.  We encountered one such argument above in the
context of cluster cores, where dark matter was detected via lensing
even though the characteristic Newtonian acceleration of the system
put it outside of the MOND regime.  

A second argument has been offered by Hoekstra, Yee \& Gladders
(2002), who have looked at the statistical galaxy-galaxy lensing
signal in the SLOAN survey.  They find that the signal is compatible
with massive halos, but more importantly that the halos must be
elliptical and aligned with the lensing galaxies.  This is significant
because far from an isolated galaxy the Newtonian potential -- and
hence the MOND lensing signature -- should arguably be highly
spherical.  This is a problem for MOND, but I can think of three ways
that non-spherical lensing could occur around galaxies in MOND. First,
the external field effect should be important at such large radii, and
would lead to a non-spherical MOND potential.  However, as pointed out
by Hoekstra et al., it does not empirically appear to be the case that
galaxies are aligned with large-scale structure (and hence the
external field).  Second, baryonic filaments in which the galaxies are
embedded could conceivably add an anisotropic lensing signal.  Again,
the difficulty would be in accounting for the alignment.  Finally, if
dark matter such as neutrinos is required in MOND, galaxies might be
embedded in rather distended neutrino halos that, if elliptical, could
add to the lensing signal just as in CDM.  However, this is nothing
more than speculation. In short, I would judge that the data of
Hoekstra et al. provide a strong, though perhaps not ironclad,
argument against MOND.

A final and significant worry for MOND is the recent accumulated data
on flux ratio anomalies in strong lensing of quasars by galaxies.  It
appears now to be widely agreed that -- at least for radio
observations -- these flux ratios cannot result from any reasonable
lens model based on smoothly distributed matter looking anything like
the distribution of visible matter.  This has been taken as evidence
for dark matter substructure in galaxies (e.g., Dalal \& Kochanek
2002).  While the issue has not (to my knowledge) been explicitly
discussed in the literature, I have an extremely hard time seeing how
this data could be accounted for in MOND, given that the lensing
occurs in the Newtonian (not MOND) acceleration regime, and that
in MOND the required substructure simply should not be there. This
argument, if it cannot be somehow circumvented, would similarly appear
to doom other modified gravity theories.

\section{MOND and the CMB}

As in the case of lensing, MOND cannot be used to make a rigorous
prediction regarding the CMB anisotropies, because there is no relativistic
cosmological framework in which they can be calculated. Nonetheless,
as for lensing there are several extant arguments concerning MOND and
the CMB.

The first argument proceeds as follows.  Plot the WMAP power spectrum
of CMB anisotropies, and overlay a predicted power spectrum from a
standard $\Lambda$CDM model.  Even those most skeptical of standard
cosmology must admit that the agreement is amazing.  It is hard to see
how a theory with significantly different physics at the recombination
epoch could give a similar agreement while still fitting data
concerning the large-scale structure growing from the same
perturbations.  The counter-argument that this agreement requires the
introduction of not one but {\em two} forms of completely mysterious
and unidentified stuff (dark matter and dark energy) is compelling,
but substantially weakened by the completely independent astrophysical
evidence pointing to -- and quantifying both of these mysterious
components.  While this sort of ``wow'' argument does not, logically
speaking, say anything whatever about MOND, it goes a long way toward
convincing many cosmologists that the standard model is on the right
track, and (unless Nature is very cruel) hence that any other track
must be incorrect.

The second argument for testing MOND using the CMB is to assert that
whatever the relativistic generalization of MOND, it will 
have the features that a) the acceleration scale $a_0$ is fixed in
time, b) the relevant acceleration for determining whether dynamics
are modified is the peculiar acceleration (and not that of the cosmic
expansion).  It can then be argued that at very early times, at all
relevant length scales the characteristic accelerations of density
perturbations are $> a_0$, and hence evolve exactly as in standard
gravity (e.g., McGaugh 1999). This means that the physics of the CMB
should be substantially unchanged except for the absence of cold dark
matter, and a possible change in the angular diameter distance due to
MONDifications of dynamics at low-$z$.  This provides a strong test
for MOND, because without collisionless, cold dark matter the power
spectrum peak heights decrease monotonically as specified by the
damping term due (primarily) to the finite thickness of the last
scattering surface and photon diffusion.  Thus observation of
alternating peak heights would be extremely difficult to account for
in MOND (or other alternatives to dark matter). The only
recourse is probably to assume that gravity {\em is} modified at the
recombination epoch, but then the ``wow'' argument of the success of
the standard scenario returns in force.  The present CMB data suggests
a high third peak, but the reader will have to judge for themselves
whether the data is strong enough to doom MOND on this count.

Even sans a third peak MOND is by no means safe.  It was asserted by
David Spergel at this conference that the first two peaks alone
contain ample data to rule out possibilities very different from the
standard parameter values, for example, a $\Omega_b=0.04$,
$\Omega_\Lambda=0.96$ universe.  On the other hand, in MOND one might
have the freedom to rescale the angular diameter distance, and to
include massive neutrinos (which would become non-relativistic near
the recombination epoch).  No detailed study has been performed of
this question, and it would be interesting to know if there is any
good fit to the WMAP data in the allowed MOND parameter
space.\footnote{S. McGaugh has informed me that he can obtain an
acceptable fit to the WMAP data with no dark matter, a reasonable
$\Omega_b$, and $\Omega_\nu \sim (1-1.5)\Omega_b$ in heavy neutrinos.}

\section{Summary: Tests of Alternatives to Dark Matter}
\begin{table}
\caption{Summary of tests of alternatives to dark matter.}
\begin{tabular}{|l|l|l|}
\tableline
{\bf Test/Prediction} & {\bf MOND} & {\bf DM} \\
\tableline
Correct rotation curve shapes & $\surd\surd$ & $\surd\times$ \\
Correct T-F slope and intercept & $\surd\surd$ & $\times$ \\
Visible matter $\rightarrow$ rotation curves & $\surd\surd\surd$ & $\times$ \\
Elliptical galaxy properties & $\surd$? & $\surd$ \\
Cluster temperature profiles & $\times$ & $\surd$ \\
CMB spectrum shapes (inc. alternating peaks) & $\times$? & $\surd\surd\surd$ \\
Correct lensing & $\times$ & $\surd\surd$ \\
Correct large-scale structure, BBN & $?$ & $\surd\surd\surd$ \\
\tableline
\end{tabular}
\end{table}

The wonderful thing about physical theories is that they are testable,
and good theories give ample specific predictions that are capable of
falsification.  I think it can be argued that for a long time CDM did
{\em not} do this, and led to no small discomfort among more skeptical
parties.  But the situation has clearly changed, and CDM has
passed important and stringent tests, especially on cosmological
scales.  At galactic scales the situation is less clear, and much of
this conference concerned the question of to what level current
observations conflict with the theory of CDM, given that making those
predictions is presently very difficult.

Just as for dark matter, there are a suite of tests for alternatives.
I have summarized a number of these in the table, and have focused on
a comparison between MOND and CDM.  The table is provided largely
without detailed justification as an expression of my opinion, with check marks
indicating (in my judgment) successes and crosses signifying what I consider to
be difficulties.  

Here are some key points, however:

\noindent 1. The success of MOND/MIFF in galaxies strongly suggests any theory
(modified gravity or dark matter) than cannot reproduce MONDian behavior
in spiral galaxies is in trouble.

\noindent 2. MOND makes no firm prediction for relativistic phenomena but at
least has a possible way to avoid altering the (well tested) physics
of the early universe drastically.  This may not be the case for other
modified gravity theories that do make, in principle, rigorous
predictions for, e.g. nucleosynthesis and CMB anisotropies which ought
to disagree with the standard analysis.

\noindent 3. Nearly independent of details, the shape of the CMB
anisotropy spectrum and evidence from lensing for unseen substructure
in galaxies constitute grave challenges for MOND or other dark matter
alternatives.

\section{Unanswered Questions}

I would like to conclude with a list of questions to which I would
like to know the answers and which suggest to me interesting avenues
of future research regarding the issues of galaxy formation, dark
matter,and modified gravity.

{\noindent\em 1. Does a satisfactory modified relativistic gravity
(MORG?) exist?}  The no-go theorems suggest this will not be a simple
extension of GR.  Milgrom has done interesting work exploring MOND as
a modification of inertia; the hope would be to make an explicit link
to cosmology via Mach's principle and to explain the coincidence
between $a_0$ and $c\sqrt{\Lambda}$ or $cH_0$ (which are all
numerically similar).

{\noindent\em 2. Can MOND, by hook or crook, wiggle out of its
difficulties with clusters, the CMB, and lensing?}  

Neutrinos seem a good bet for explaining clusters, but may be ruled
out soon.  The CMB constraint awaits a careful study, but it does not
seem impossible to me that a fit could be found if extra ingredients
are added.  I can offer no good ideas for explaining the lensing flux
anomalies in MOND.

{\noindent\em 3. How well does MIFF really work in galaxies?}

Does the extraordinary success of Milgrom's Fitting Formula (MIFF) in
spirals extend to ellipticals of various masses, or irregular and
dwarf galaxies?  Also, how much of the regularity arises from the
selection of very ``clean'' galaxies to analyze?  It would be good to
know, independent of MOND, how well MIFF works so that it could be
employed like the TF and fundamental plane relations.

{\noindent\em 4. If MIFF works as well as it presently appears to, why?}

Assuming that MIFF's success is not an artifact selecting particular
galaxies, it implies that there is an enormous regularity to galaxy
formation, and in particular in the correspondence between visible and
dark matter.  It is unclear to me how this great regularity arises in
realistic CDM galaxy formation scenarios which include and require a
number of stochastic components such as strong feedback and mergers.

{\noindent\em 5. If MIFF holds, but not MOND, whence $a_0$?  Does $a_0
\rightarrow$ (Dynamics), or does (Dynamics) $\rightarrow a_0$?}

The success of MIFF, and the form of other attempts at modified
gravity, strongly suggest a characteristic acceleration scale in
galaxy formation.  Where does this arise from?  There have been a
couple of stabs at this question (e.g., Kaplinghat \& Turner 2002),
but I find them rather unconvincing.  In addition, as pointed out by
Milgrom (2002), $a_0$ plays several somewhat independent roles in
MOND/MIFF.  Why so, if the near-universal acceleration at the onset of
dark-matter domination is just coincidental?  (Here, testing MIFF in a
variety of object to determine the acuteness of this question is
important.) There is a fundamental question of logical priority to
address, I believe: is $a_0$ simply an emergent number that arises
through the complexity of galaxy formation and just happens to be
nearly-universal in any phenomenologically viable galaxy formation
scenario, or is $a_0$ associated with some specific physical effect
(such as a surface density threshold) that {\em leads} to
regularities in galaxy formation -- such as the particular slope, and
surface-brightness independence of the TF relation -- that we observe?


\end{document}